\documentclass[12pt]{article}
\pdfoutput=1
\usepackage[T1]{fontenc}
\usepackage[utf8]{inputenc}
\usepackage[english]{babel}
\usepackage{amsmath}
\usepackage{amssymb}
\usepackage{mathtools} 
\usepackage{empheq} 
\usepackage{diagbox} 
\usepackage{lmodern}
\usepackage{exscale}
\usepackage{icomma}
\usepackage[section]{placeins}
\usepackage{graphicx}
\usepackage{lmodern}
\usepackage{gensymb}
\usepackage{graphicx}
\usepackage{sidecap}
\usepackage{array}
\usepackage{float}
\usepackage{dsfont}
\usepackage[hidelinks]{hyperref}
\usepackage{tabto}
\usepackage{listings}
\usepackage{xcolor}
\usepackage{fancyhdr}
\usepackage{textcomp}
\usepackage[export]{adjustbox}
\usepackage[skip=2pt,font=footnotesize]{caption}
\usepackage{physics}
\usepackage{verbatim}
\usepackage[]{pdfpages}
\usepackage[disable]{todonotes} 
\usepackage{yhmath}
\usepackage{marvosym}
\usepackage{tensor}
\usepackage{empheq}

\usepackage[portrait,paper=a4paper,hmarginratio=1:1]{geometry}
\usepackage{multirow}
\usepackage{multicol}
\usepackage{blindtext}
\usepackage[separate-uncertainty=true]{siunitx}
\usepackage{tikz}
\usetikzlibrary{shapes,arrows,decorations.pathmorphing,automata,positioning,calc,tikzmark,fit,patterns,patterns.meta}
\usepackage{tkz-euclide}
\usetikzlibrary{math}

\usepackage[sorting=none,url=false,maxbibnames=99,maxcitenames=2,giveninits=true]{biblatex}
\DeclareNameAlias{author}{last-first}
\DeclareNameAlias{editor}{last-first}
\usepackage{pgfplots}

\usepackage{relsize} 
\tikzset{fontscale/.style = {font=\relsize{#1}}
    }

\usepackage{stackrel}

\usepackage{graphics} 

\usepackage[yyyymmdd]{datetime}

\usepackage{subcaption} 

\usepackage{comment}

\usepackage{lipsum}

\usepackage{tcolorbox}

\usepackage{algorithm}
\usepackage{algpseudocode}

\usepackage{stmaryrd}

\usepackage{authblk}

\usepackage{lastpage}

\DeclareSIUnit \cycle {cycle}
\DeclareSIUnit \MPam {(\mega\pascal\sqrt\meter)}
\DeclareSIUnit \Pam {(\pascal\sqrt\meter)}

\title{Electrode and electroactive polymer layout design using topology optimization}

\date{}

\author[1,*]{Daniel Hård} 

\author[1]{Mathias Wallin} 
\author[1]{Matti Ristinmaa} 

\affil[1]{\small Division of Solid Mechanics, Lund University, Box 118, SE-22100 Lund, Sweden}
\affil[*]{\small Corresponding author: Daniel Hård, daniel.hard@solid.lth.se}

\begin{document}

\maketitle

\begin{abstract}
	\noindent When electrically stimulated, electroactive polymers (EAPs) respond with mechanical deformation. The goal of this work is to design electrode and EAP layouts simultaneously in structures by using density-based, multi-material topology optimization. In this novel approach the layout of electrodes and EAP material are not given a priori but is a result from the topology optimization. Material interpolation based on exponential functions is introduced, allowing a large flexibility to control the material interpolation. The electric field in the surrounding free space is modeled using a truncated extended domain method. Numerical examples that demonstrates the method's ability to design arbitrary EAP and electrode layouts are presented. In these optimized structures, electrode material is continuously connected from the electrical sources to opposite sides of the EAP material and thereby concentrating the electric field to the EAP material which drives the deformation.
	\\ \\
	\textit{Keywords}: Multi-material, Topology optimization, Electroactive polymer, Electrode, Free space
\end{abstract}

\section{Introduction}

Electro-mechanically coupled materials deform when electrically stimulated. One example is piezoelectric materials where electric field and deformation are linearly coupled. Another example is electroactive polymers (EAPs) which has a non-linear coupling. Dielectric EAPs, which is the subject of this study, is an active research area which has great potential in, e.g., artificial muscles and actuators \cite{bar-cohen_electroactive_2004,bar-cohen_electroactive_2019}. Typically, dielectric EAPs are constructed as thin layer of EAP material sandwiched between two compliant electrodes such that a potential difference will cause the EAP to contract in the direction of the electric field.

Modeling and design of EAPs has attracted much attention, see, e.g., \textcite{dorfmann_nonlinear_2005}, \textcite{ask_phenomenological_2012,ask_electrostriction_2012,ask_modelling_2015}, \textcite{ortigosa_new_2016}. In recent years, there has been a growing interest into combining EAPs and topology optimization. \textcite{bortot_topology_2018} used topology optimization with a genetic algorithm to design dielectric elastomer structures with tunable band gaps. This problem was later solved more efficiently using gradient-based optimization by \textcite{sharma_gradient-based_2022}. \textcite{ortigosa_density-based_2021} used topology optimization to optimize the active EAP layer. This was developed into a multi-resolution model by \textcite{ortigosa_multi-resolution_2021}. In \textcite{martinez-frutos_-silico_2021} the electrode placement was optimized, however they only considered the electrodes as a 2D-layer of nodes, not as a separate material. Recently, electrode connectivity was investigated by \textcite{hard_connectivity_2024}, where the shape of the electrodes were implicitly given by the solid material phase.

The goal in this work is to develop a topology optimization methodology capable of generating arbitrary EAP and electrode layouts. With this approach, the layout of electrode and EAP material is not given a priori but is an outcome of the topology optimization. This requires the use of multi-material topology optimization.

Multi-material topology optimization can be used to find the optimal distribution of multiple different material phases. The commonly used SIMP method, see \textcite{bendsoe_material_1999}, can be extended to handle multiple materials. Alternative methods include, e.g., Discrete Material Optimization (DMO) by \textcite{stegmann_discrete_2005} and Unified Material Interpolation (UMI) by \textcite{yi_unified_2023}. In the context of multi-material optimization of coupled problems, \textcite{granlund_topology_2024} considered transient thermo-mechanical problems and \textcite{he_multi-material_2021} considered piezoelectric structures for energy harvesting. \textcite{hu_integrated_2024} investigated multi-material and multi-scale topology optimization of structures with embedded piezoelectric actuators. \textcite{de_almeida_topology_2024} proposed a method for simultaneous optimization of dielectric and conductive materials together with a geometry projection method for embedded piezoelectric stack actuators. \textcite{ortigosa_programming_2023} investigated multi-material shape morphing of layered EAPs, but did not optimize the electrodes explicitly.

In most research on EAPs, only solid bodies were modeled. This is a reasonable approach for materials with large permittivity, e.g., piezoelectric materials. In the present case of dielectric EAPs, the permittivity is significantly smaller and approximately of the same or one order larger than vacuum. This will have substantial influence on the modeling of the EAP. Multiple approaches to model the electric fields in the surrounding free space for electrostatic conditions exist, see, e.g., \textcite{chen_review_1997}. \textcite{vu_2-d_2010,vu_3-d_2012} used a coupled BEM-FEM method to model the electric field in the space surrounding an EAP. \textcite{pelteret_computational_2016} used a truncated free space in the study of a mixed variational formulation for quasi-incompressible electroactive and magnetoactive polymers. \textcite{dev_influence_2023} incorporated a truncated free space in the context of topology optimization and this methodology is adapted in the current work.

The structure of this paper is as follows. In Section \ref{sec:EAP} the EAP model is described. In Section \ref{sec:TopOpt}, the topology optimization formulation is developed. Numerical examples that demonstrates the method's ability to generate arbitrary EAP and electrode layouts are shown in Section \ref{sec:NumericalEx}. Lastly, Section \ref{sec:Conclusions} provides conclusions.

\section{Non-linear electro-elasticity with free space}
\label{sec:EAP}
In this section, the governing non-linear electro-elastic balance laws restricted to electrostatic condition will be introduced. For a detailed presentation of the topic, the reader is referred to \textcite{jackson_classical_1962}, \textcite{eringen_electrodynamics_1990}, \textcite{kovetz_electromagnetic_2000}, and \textcite{dorfmann_nonlinear_2005}. Modeling the surrounding free space is especially important for the electric field, since it is not restricted to the solid body. This work models the free space as a truncated extended domain presented in \textcite{dev_influence_2023}, where the free space is modeled as a very soft elastic material with vacuum permittivity properties.

\begin{figure}[H]
	\centering
	\includegraphics{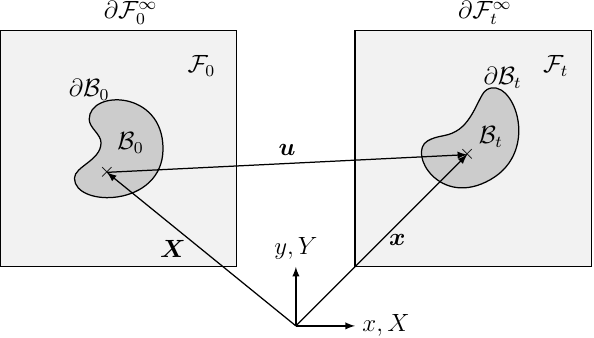}
	\caption{A solid body $\mathcal{B}_0$ surrounded by a free space $\mathcal{F}_0$ in the material and spatial configurations.}
	\label{fig:FreeSpace}
\end{figure}

Fig. \ref{fig:FreeSpace} illustrates the considered domain in both the material and spatial configurations. The domain consists in the material configuration of a body $\mathcal{B}_0$ and the surrounding free space $\mathcal{F}_0$ such that the undeformed domain is given by $\mathcal{B}_0\cup\mathcal{F}_0$ and $\mathcal{B}_0\cap\mathcal{F}_0=\varnothing$. When deformed, $\mathcal{B}_0$ and $\mathcal{F}_0$ are mapped to $\mathcal{B}_t$ and $\mathcal{F}_t$, respectively. Particles are labeled by their coordinates $\vb*{X}\in\mathcal{B}_0\cup\mathcal{F}_0$ and their placement $\vb*{x}\in\mathcal{B}_t\cup\mathcal{F}_t$ is described by the smooth mapping $\vb*{x}(\vb*{X})=\vb*{\varphi}(\vb*{X})$. The deformation is locally described by the deformation gradient $\vb*{F}=\nabla_{\vb*{X}}\vb*{\varphi}$ and the local volume change $J=\det(\vb*{F})$. The hyper-elastic material response will be governed by the right Cauchy-Green tensor defined as $\vb*{C} =\vb*{F}^T\cdot\vb*{F}$ and its isochoric part $\overline{\vb*{C}}=J^{-2/3}\vb*{C}$.

\subsection{Balance relations}

Assuming quasi-electrostatics, the Maxwell-Faraday equation becomes $\nabla_{\vb*{X}}\cross\vb*{E} = \vb*{0}$, implying that the electric field $\vb*{E}$ is obtained from the gradient of a continuous scalar electric potential $\phi$ as $\vb*{E} = -\nabla_{\vb*{X}}\phi$, where the negative sign is introduced by convention.

Gauss's law, assuming no free volume charge density in the dielectric material, is formulated as
\begin{subequations}
	\begin{alignat}{2}    
	\nabla_{\vb*{X}}\cdot\vb*{D} &= 0, & \quad&  \text{in } \mathcal{F}_0\cup\mathcal{B}_0,
	\label{eq:SpatGaussLaw}
	\\
	\phi &= \phi_g, & & \text{on } \partial\mathcal{B}_0^{g,\phi},
	\label{eq:electricPotBC}
	\\
	\vb*{D}\cdot\vb*{N} &= 0, & & \text{on } \partial\mathcal{F}_0^{\infty},
	\label{eq:MatGaussLawBCinfty}
	\\
	-\llbracket\vb*{D}\rrbracket\cdot\vb*{N} &= D_N,& & \text{on } \partial\mathcal{B}_0^{h,\phi},
	\label{eq:MatGaussLawBC}
	\end{alignat}
\end{subequations}
where $\vb*{D}$ is the electric displacement. The boundary $\partial\mathcal{B}_0$ to $\mathcal{B}_0$ is split such that $\partial\mathcal{B}_0=\partial\mathcal{B}_0^{h,\phi}\cup\partial\mathcal{B}_0^{g,\phi}$ and $\partial\mathcal{B}_0^{h,\phi}\cap\partial\mathcal{B}_0^{g,\phi}=\varnothing$. The boundary condition \eqref{eq:electricPotBC} represents the prescribed electric potential on the boundary part $\partial\mathcal{B}_0^{g,\phi}$, and the boundary condition \eqref{eq:MatGaussLawBC} represents the prescribed free surface charge density on the boundary part $\partial\mathcal{B}_0^{h,\phi}$, with the outward unit normal $\vb*{N}$ to the body $\mathcal{B}_0$. The jump $\llbracket \bullet \rrbracket = [\bullet]^+ -[\bullet]^- $ is the difference in the value $(\bullet)$ between inside and outside the body. The condition in \eqref{eq:MatGaussLawBCinfty} on the far-field boundary $\partial\mathcal{F}_0^\infty$ ensures that the electric field becomes zero at large distances from the solid body.

In the absence of body forces, the balance of linear momentum for quasi-static electro-mechanics is provided by
\begin{subequations}
	\begin{alignat}{2}
	\nabla_{\vb*{X}}\cdot\vb*{T}
	&= \vb*{0},& \quad &\text{in } \mathcal{F}_0\cup\mathcal{B}_0,
	\label{eq:SpatTotalBalanceOfMomentum}
	\\
	\vb*{\varphi} &=\vb*{\varphi}_g, & & \text{on } \partial\mathcal{B}_0^{g,\vb*{\varphi}},
	\label{eq:SpatTotalBalanceOfMomentumDirichletBC} 
	\\
	\vb*{\varphi} &= \vb{0}, & & \text{on } \partial\mathcal{F}_0^{\infty},
	\label{eq:SpatTotalBalanceOfMomentumDirichletBCinfty}
	\\
	\llbracket\vb*{T}\rrbracket\cdot\vb*{N} &= \vb*{T}_N,& &\text{on } \partial\mathcal{B}_0^{h,\vb*{\varphi}},
	\label{eq:SpatTotalBalanceOfMomentumNeumannBC}  
	\end{alignat}
\end{subequations}
where $\vb*{T}$ is the total first Piola-type stress tensor. The boundary $\partial\mathcal{B}_0$ to $\mathcal{B}_0$ is split such that $\partial\mathcal{B}_0=\partial\mathcal{B}_0^{h,\vb*{\varphi}}\cup\partial\mathcal{B}_0^{g,\vb*{\varphi}}$ and $\partial\mathcal{B}_0^{h,\vb*{\varphi}}\cap\partial\mathcal{B}_0^{g,\vb*{\varphi}}=\varnothing$. The boundary condition in \eqref{eq:SpatTotalBalanceOfMomentumDirichletBC} represents the prescribed displacement on $\partial\mathcal{B}_0^{g,\vb*{\varphi}}$, \eqref{eq:SpatTotalBalanceOfMomentumDirichletBCinfty} represents the fixed displacement at large distances, and \eqref{eq:SpatTotalBalanceOfMomentumNeumannBC} represents the prescribed total traction on $\partial\mathcal{B}_0^{h,\vb*{\varphi}}$. It is assumed that the free space can be modeled with \eqref{eq:SpatTotalBalanceOfMomentum}.

\subsection{Constitutive model}
The constitutive dielectric EAP model used in this work can be found in, e.g., \textcite{ask_phenomenological_2012,ask_electrostriction_2012,ask_modelling_2015}, in this work the viscoelasticity contribution is omitted. For simplicity, the same model is used to model the electrode material. An augmented free energy function $\Omega\qty(\vb*{F},\vb*{E})$ can be introduced, see, e.g., \textcite{dorfmann_nonlinear_2005}, which defines $\vb*{D}$ and $\vb*{T}$, i.e.,
\begin{equation}
\vb*{D}(\vb*{F},\vb*{E})
= -\pdv{\Omega\qty(\vb*{F},\vb*{E})}{\vb*{E}},
\quad
\vb*{T}(\vb*{F},\vb*{E}) 
= \pdv{\Omega\qty(\vb*{F},\vb*{E})}{\vb*{F}},
\label{eq:augmConst}
\end{equation}
where the augmented free energy function is taken as
\begin{equation}
\begin{split}
\Omega(\vb*{F},\vb*{E}) =& \frac{1}{2}K\qty(J-1)^2 + \frac{1}{2}G\qty(\overline{\vb*{C}}:\vb*{I}-3)
\\
&-\frac{1}{2}\varepsilon_0\varepsilon_r J\vb*{E} \cdot\vb*{C}^{-1} \cdot\vb*{E} + c_e\vb*{E}\cdot\vb*{E},
\end{split}
\label{eq:AugmentedFreeEnergy}
\end{equation} 
where $K$ and $G$ are the bulk and shear modulus, respectively, $\varepsilon_0\approx\SI{8.854e-12}{\farad\per\metre}$ the vacuum permittivity, $\varepsilon_r$ is the relative permittivity, and $c_e$ an electrical material constant.

Inserting \eqref{eq:AugmentedFreeEnergy} in \eqref{eq:augmConst} yields
\begin{equation}
\vb*{D}(\vb*{F},\vb*{E})
= \varepsilon_0\varepsilon_r J\vb*{E}\cdot\vb*{C}^{-1} - 2c_e\vb*{E},
\label{eq:explicitDfromOmega}
\end{equation}
and
\begin{equation}
\begin{split}
\vb*{T}(\vb*{F},\vb*{E}) 
=& K\qty(J-1)J\vb*{F}^{-T} 
+ G J^{-2/3}\bigg[\vb*{F} - \frac{1}{3}[\vb*{C}:\vb*{I}]\vb*{F}^{-T}\bigg] 
\\
&+ \varepsilon_0\varepsilon_r J\bigg[\qty(\vb*{E}\cdot\vb*{F}^{-1})\otimes\qty(\vb*{E}\cdot\vb*{C}^{-1})
-\frac{1}{2}(\vb*{E}\cdot\vb*{C}^{-1}\cdot\vb*{E})\vb*{F}^{-T}\bigg].
\end{split}
\label{eq:explicitTfromOmega}
\end{equation}

The free space is modeled as void material in the topology optimization introduced in Section \ref{sec:TopOpt}. It is well known that the Neo-Hookean approach in \eqref{eq:AugmentedFreeEnergy} can introduce instabilities, which could be resolved with a linear elasticity model, e.g. by gamma scaling as in \textcite{wang_interpolation_2014}. This was however not necessary in this work. Additional problems noted in \textcite{ortigosa_density-based_2021} is that the electro-mechanical term in \eqref{eq:AugmentedFreeEnergy} may result in numerical instabilities due to high non-convexity. Therefore, the stabilized formulation proposed by \textcite{ortigosa_density-based_2021} is used instead, which removes the dependency on the deformation gradient. The same approach was also used in \textcite{dev_influence_2023}. The free energy for the void region is taken as
\begin{align}
\begin{split}
\Omega^\text{void}(\vb*{F},\vb*{E})
=&
\frac{1}{2}K^\text{void}\qty(J-1)^2 
+ \frac{1}{2}G^\text{void}\qty(\overline{\vb*{C}}:\vb*{I}-3)
\\ 
&- \frac{1}{2} \varepsilon_0\vb*{E}\cdot\vb*{E},
\end{split}
\label{eq:AugementedFreeEnergyVoid}
\end{align}
where $K^\text{void}$ and $G^\text{void}$ are sufficiently small modeling parameters to not significantly influence the mechanical response while large enough to avoid numerical convergence issues.

\subsection{Variational forms and finite element discretization}
The weak forms of Gauss's law \eqref{eq:SpatGaussLaw} and the electro-mechanical balance of linear momentum \eqref{eq:SpatTotalBalanceOfMomentum} in the material configuration including both the body $\mathcal{B}_0$ and the free space $\mathcal{F}_0$ are
\begin{subequations}
	\begin{align}
	\delta W^{\text{elt}}
	&= \delta W^{\text{elt}}_{\text{ext}} 
	- \delta W^{\text{elt}}_{\text{int}}
	= 0,\quad 
	& &\forall \delta\phi \in  \qty[v|v=0\text{ on }\partial\mathcal{B}_0^{g,\phi}],
	\label{eq:virtWorkElt}
	\\
	\delta W^{\text{mec}}
	&= \delta W^{\text{mec}}_{\text{ext}} 
	- \delta W^{\text{mec}}_{\text{int}}
	= 0, 
	& &\forall\delta\vb*{\varphi}  \in   \qty[\vb*{v}|\vb*{v}=\vb*{0}\text{ on }\partial\mathcal{B}_0^{g,\vb*{\varphi}}\cup\partial\mathcal{F}_0^{\infty}],
	\label{eq:virtWorkMech}
	\end{align}
\end{subequations}
where
\begin{subequations}
	\begin{align}
	\delta W^{\text{elt}}_{\text{ext}}
	&= -\int_{\partial\mathcal{B}_0^{h,\phi}}\delta\phi D_N \dd A,
	\\
	\delta W^{\text{elt}}_{\text{int}}
	&=  \int_{\mathcal{F}_0\cup\mathcal{B}_0} \nabla_{\vb*{X}}\delta\phi\cdot\vb*{D}\dd V,
	\\
	\delta W^{\text{mec}}_{\text{ext}}
	&= \int_{\partial\mathcal{B}_0^{h,\vb*{\varphi}}}\delta\vb*{\varphi}\cdot \vb*{T}_N \dd A, 
	\\
	\delta W^{\text{mec}}_{\text{int}}
	&=\int_{\mathcal{F}_0\cup\mathcal{B}_0} \nabla_{\vb*{X}}\delta\vb*{\varphi}:\vb*{T}\dd V.
	\end{align}
\end{subequations}

A standard nonlinear finite element formulation is used to discretize \eqref{eq:virtWorkElt} and \eqref{eq:virtWorkMech} where the same shape functions are used to interpolate both the displacements $\vb*{\varphi}$ and potential $\phi$ as $\vb*{\varphi}\approx\vb*{N}^{\vb*{\varphi}}\vb*{a}^{\vb*{\varphi}}$ and $\phi\approx\vb*{N}^{\phi}\vb*{a}^{\phi}$, where $\vb*{a}^{\vb*{\varphi}}$ and $\vb*{a}^{\phi}$ are the nodal values and $\vb*{N}^{\vb*{\varphi}}$ and $\vb*{N}^{\phi}$ the corresponding global shape functions. Only one additional degree of freedom is needed for the potential. The virtual displacements and virtual electric potential are approximated using the Galerkin approach as $\delta\vb*{\varphi}\approx\vb*{N}^{\vb*{\varphi}}\delta\vb*{a}^{\vb*{\varphi}}$ and $\delta\phi\approx \vb*{N}^{\phi}\delta\vb*{a}^{\phi}$, where $\delta\vb*{a}^{\vb*{\varphi}}$ and $\delta\vb*{a}^{\phi}$ are the virtual nodal values.

The finite element approximations inserted into \eqref{eq:virtWorkElt} and \eqref{eq:virtWorkMech} together with the arbitrariness of $\delta\vb*{a}^{\vb*{\varphi}}$ and $\delta\vb*{a}^{\phi}$ results in the residual equations on matrix format
\begin{subequations}
	\begin{align}
	\vb*{r}^{\phi} 
	&= \vb*{f}_{ext}^{\phi} - \vb*{f}_{int}^{\phi}
	= \vb*{0},
	\label{eq:TotalLagrangianFEresidualsPot}
	\\
	\vb*{r}^{\vb*{\varphi}}
	&= \vb*{f}_{ext}^{\vb*{\varphi}} - \vb*{f}_{int}^{\vb*{\varphi}}
	= \vb*{0},
	\label{eq:TotalLagrangianFEresidualsDef} 
	\end{align}
\end{subequations}
where
\begin{subequations}
	\begin{align}
	\vb*{f}_{ext}^{\phi}
	&=\int_{\partial\mathcal{B}_0^{h,\phi}} \vb*{N}^{\phi,T}D_N \dd A,
	\label{eq:}
	\\
	\vb*{f}_{int}^{\phi} 
	&= \int_{\mathcal{F}_0\cup\mathcal{B}_0} \vb*{B}^{\phi,T}\vb*{D} \dd V,
	\label{eq:fintPot}
	\\
	\vb*{f}_{ext}^{\vb*{\varphi}}
	&=\int_{\partial\mathcal{B}_0^{h,\vb*{\varphi}}} \vb*{N}^{\vb*{\varphi},T}\vb*{T}_N \dd A,
	\\
	\vb*{f}_{int}^{\vb*{\varphi}}
	&= \int_{\mathcal{F}_0\cup\mathcal{B}_0} \vb*{B}^{\vb*{\varphi},T}\vb*{T} \dd V,
	\label{eq:fintDef}
	\end{align}
\end{subequations}
where $\vb*{B}^{\phi}$ and $\vb*{B}^{\vb*{\varphi}}$ contains the gradients of $\vb*{N}^{\phi}$ and $\vb*{N}^{\vb*{\varphi}}$.

Newton's method is used to solve \eqref{eq:TotalLagrangianFEresidualsPot} and \eqref{eq:TotalLagrangianFEresidualsDef}. Linearization and assuming dead loading condition yields a discretized and linearized system, which in monolithic matrix format it is formulated as
\begin{equation}
\vb*{K}\Delta\vb*{a} =
\mqty[\vb*{K}^{\vb*{\varphi}\vb*{\varphi}} & \vb*{K}^{\vb*{\varphi}\phi}\\\vb*{K}^{\phi\vb*{\varphi}} & \vb*{K}^{\phi\phi}]
\mqty[\Delta\vb*{a}^{\vb*{\varphi}}\\\Delta\vb*{a}^\phi]
= \mqty[\vb*{r}^{\vb*{\varphi}}\\\vb*{r}^\phi]
=\vb*{r},
\label{eq:LinarisedSystem}
\end{equation}
which provides the Newton updates
\begin{equation}
\vb*{a}^{\vb*{\varphi}} \leftarrow \vb*{a}^{\vb*{\varphi}} + \Delta\vb*{a}^{\vb*{\varphi}},
\quad 
\vb*{a}^{\phi} \leftarrow \vb*{a}^{\phi} + \Delta\vb*{a}^{\phi},
\label{eq:NewtonUpdate}
\end{equation}
and continues until the residuals in \eqref{eq:TotalLagrangianFEresidualsPot} and \eqref{eq:TotalLagrangianFEresidualsDef} are sufficiently small.

The partitions of $\vb*{K}$ in \eqref{eq:LinarisedSystem} are obtained from linarizations of \eqref{eq:fintPot} and \eqref{eq:fintDef}, i.e.,
\begin{align}
\begin{split}
\vb*{K}^{\phi\phi} 
&= \int_{\mathcal{F}_0\cup\mathcal{B}_0} \vb*{B}^{\phi,T} \vb*{D}^{\text{elt}} \vb*{B}^{\phi}\dd V,
\\
\vb*{K}^{\vb*{\varphi}\phi^T} = \vb*{K}^{\phi\vb*{\varphi}} 
&= -\int_{\mathcal{F}_0\cup\mathcal{B}_0} \vb*{B}^{\phi,T} \vb*{D}^{\text{mix}}\vb*{B}^{\vb*{\varphi}}\dd V,
\\
\vb*{K}^{\vb*{\varphi}\vb*{\varphi}} 
&= \int_{\mathcal{F}_0\cup\mathcal{B}_0} \vb*{B}^{\vb*{\varphi},T} \vb*{D}^{\text{mec}}\vb*{B}^{\vb*{\varphi}}  \dd V,
\end{split}
\label{eq:UpdateLagrangianFE} 
\end{align}
where $\vb*{D}^{\text{elt}}$, $\vb*{D}^{\text{mix}}$ and $\vb*{D}^{\text{mec}}$ are the Voigt representation of the material tangents,
\begin{alignat}{3}
\boldsymbol{\mathsf{D}}^{\text{elt}}&= \pdv{\Omega}{\vb*{E}\otimes}{\vb*{E}},
\quad
&\boldsymbol{\mathsf{D}}^{\text{mix}} 
&= \pdv{\Omega}{\vb*{E}\otimes}{\vb*{F}},
\quad
&\boldsymbol{\mathsf{D}}^{\text{mec}} &= \pdv{\Omega}{\vb*{F}\otimes}{\vb*{F}},
\label{eq:Tangents}
\end{alignat}
where $\boldsymbol{\mathsf{D}}^{\text{elt}}$ is second order, $\boldsymbol{\mathsf{D}}^{\text{mix}}$ third order, and $\boldsymbol{\mathsf{D}}^{\text{mec}}$ fourth order tensors. 

\section{Topology optimization}
\label{sec:TopOpt}

The design, in the undeformed body domain $\mathcal{B}_{0}$, is defined by the two piecewise uniform density fields $\rho_1$ and $\rho_2$. Each finite element $e$ in $\mathcal{B}_{0}$ is associated with two uniform element densities $\rho_i^e\in[0,1]$ which are collected in the design vectors $\vb*{\rho}_1$ and $\vb*{\rho}_2$. The density field $\rho_1$ controls the amount of solid material and $\rho_2$ if the solid material is electrode or EAP. This is further explained in Section \ref{sec:MaterialInterpolation}.

Using a nested format, the general form of the optimization problem is formulated as
\begin{equation}
(\mathbb{TO})
\begin{cases}
\displaystyle \min_{\vb*{\rho}_1,\vb*{\rho}_2} g_0 \\
\displaystyle \text{subjected to}
\begin{cases}
\displaystyle g_j \leq 0,\quad j=1,...,n_{constr}\\
\displaystyle 0\leq \rho_i^e\leq1,\quad \forall e\in[1,...,n_{elm}], i\in[1,2]
\end{cases}
\end{cases}
\label{eq:SO}
\end{equation}
The objective is to maximize the displacements, i.e.,
\begin{equation}
g_0(\vb*{a})=\vb*{l}^T\vb*{a},    
\label{eq:g0maxDisp}
\end{equation}
where $\vb*{l}$ is a constant vector with nonzero entries in the dofs which should be maximized, $\vb*{a}$ is the solution vector at the terminal load step.

To limit the volume of each material phase, volume constraints are enforced in \eqref{eq:SO} as,
\begin{equation}
g_1=\frac{V_1}{\alpha_1 V_{DD}}-1\leq 0,\quad
g_2=\frac{V_2}{\alpha_2 V_{DD}}-1\leq 0,
\label{eq:volumeConstraint}
\end{equation}
where $\alpha_1$ and $\alpha_2$ are the allowed volume fractions and $V_{DD}=\text{vol}(\mathcal{B}_{0})$ is the volume of the design domain. In \eqref{eq:volumeConstraint}, the volumes $V_1$ and $V_2$ of phase 1 and phase 2 are computed as
\begin{equation}
V_1
= \int_{\mathcal{B}_{0}} \Bar{\rho}_1(1-\Bar{\rho}_2)\  \dd V,
\quad
V_2
= \int_{\mathcal{B}_{0}} \Bar{\rho}_1\Bar{\rho}_2\  \dd V,
\label{eq:volume}
\end{equation}
where $\Bar{\rho}_1$ and $\Bar{\rho}_2$ are the filtered and thresholded densities as presented in Section \ref{sec:RegProInt}.

\subsection{Regularization, projection, and intermediate densities}
\label{sec:RegProInt}
The piecewise uniform density fields $\rho_1$ and $\rho_2$ are regularized by smoothing using the partial differential equation (PDE) filter proposed by \textcite{lazarov_filters_2011},
\begin{subequations}
	\begin{alignat}{2}
	-l_i^2\Delta_{\vb*{X}}\Tilde{\rho}_i + \Tilde{\rho}_i &= \rho_i, &\quad &\text{in } \mathcal{B}_{0},
	\label{eq:Filter}
	\\
	\nabla_{\vb*{X}}\Tilde{\rho}_i \cdot \vb*{N} &= 0, &\quad &\text{on } \partial\mathcal{B}_{0},
	\label{eq:FilterBC}
	\end{alignat}
\end{subequations}
where $\Delta_{\vb*{X}}$ is the material Laplacian operator, $l_i>0$ is the filter length scale parameter, and $\Tilde{\rho}_i$ the filtered, continuous fields. Note the possibility to use different length scales for $\Tilde{\rho}_1$ and $\Tilde{\rho}_2$. The boundary condition in \eqref{eq:FilterBC} tends to generate designs that artificially place material at the boundaries of $\mathcal{B}_{0}$. To reduce this, \textcite{wallin_consistent_2020} proposed using a Robin boundary condition, however this was not further investigated in this work.

Utilizing the same discretization as in the state problem render a linear system on the form,
\begin{equation}
\vb*{A}\vb*{\Tilde{\rho}}_i = \vb*{G}\vb*{\rho}_i,
\label{eq:FilterFEM}
\end{equation}
where the matrices $\vb*{A}$ and $\vb*{G}$ are design independent and only assembled once.

The amount of intermediate densities caused by the filter in \eqref{eq:Filter} is decreased by the smooth Heaviside projection introduced by \textcite{wang_projection_2011} as,
\begin{equation}
\Bar{\rho}_i 
= H_{\beta,\eta}(\Tilde{\rho}_i) 
= \frac{\tanh(\beta\eta)+\tanh(\beta(\Tilde{\rho}_i-\eta))}{\tanh(\beta\eta)+\tanh(\beta(1-\eta))},
\label{eq:Hproj}
\end{equation}
where $\beta$ controls the sharpness and $\eta$ the projection threshold. The projection \eqref{eq:Hproj} was applied on the filtered density fields $\Tilde{\rho}_i$ in every Gauss point separately for the two density fields.

\begin{figure}
	\centering
	\includegraphics{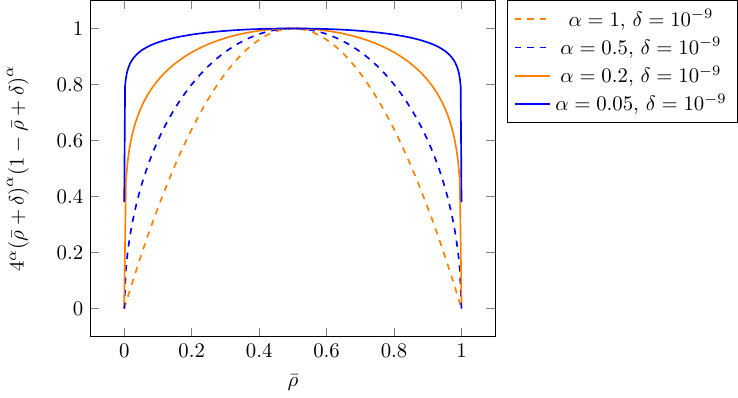}
	\caption{The penalization function in \eqref{eq:CRISP} with example values on $\alpha$ and $\delta$.}
	\label{fig:CRISP}
\end{figure}

Despite the Heaviside projection in \eqref{eq:Hproj}, the optimization might still generate designs with intermediate densities. To penalize intermediate densities, the penalization proposed in \textcite{thillaithevan_inverse_2024} and \textcite{granlund_large-scale_2024}, is adopted as,
\begin{equation}
\bar{g}_0 = g_0 + a_{d} \sum_{i=1}^2 d_\delta^\alpha(\Bar{\rho}_i),
\label{eq:addCRISPterms}
\end{equation}
where $a_{d}$ is a scale factor and
\begin{equation}
d_\delta^\alpha(\Bar{\rho}_i) = \frac{1}{V_{DD}}\int_{\mathcal{B}_{0}} 4^\alpha \qty(\bar{\rho}_i+\delta)^\alpha\qty(1-\bar{\rho}_i+\delta)^\alpha \dd V,
\label{eq:CRISP}
\end{equation}
where $V_{DD}$ normalizes the expression. In \eqref{eq:CRISP}, $0<\alpha\leq1$ and $0<\delta \ll 1$ are needed to avoid singularities at $\Bar{\rho}_i=0$ and $\Bar{\rho}_i=1$ in the sensitivities. Examples of the penalization are given in Fig. \ref{fig:CRISP}.

\subsection{Material interpolation}
\label{sec:MaterialInterpolation}

The commonly used SIMP material interpolation, see \textcite{bendsoe_material_1999}, results in that the highest sensitivities always occur for $\Bar{\rho}=1$, regardless of interpolation order. Other methods for multi-material interpolation include DMO, see \textcite{stegmann_discrete_2005}, and UMI, see \textcite{yi_unified_2023}. Both of these methods have the advantage that they are independent of interpolation order. However, numerical tests showed that in the current application, especially in combination with the penalization in \eqref{eq:CRISP}, these methods were not able to generate distinct material phases.

To remedy these problems, a formulation based on exponential functions is proposed in this work denoted as Exponential Material Interpolation (EMI),
\begin{equation}
\chi_q(\Bar{\rho}) = \frac{e^{q \bar{\rho}}-1}{e^{q}-1},
\end{equation}
where $q$ is a penalization parameter. A similar function was proposed in \textcite{ye_plateshell_2016} as a filter function, denoted composite exponential function (CEF).

\begin{figure}
	\centering
	\includegraphics{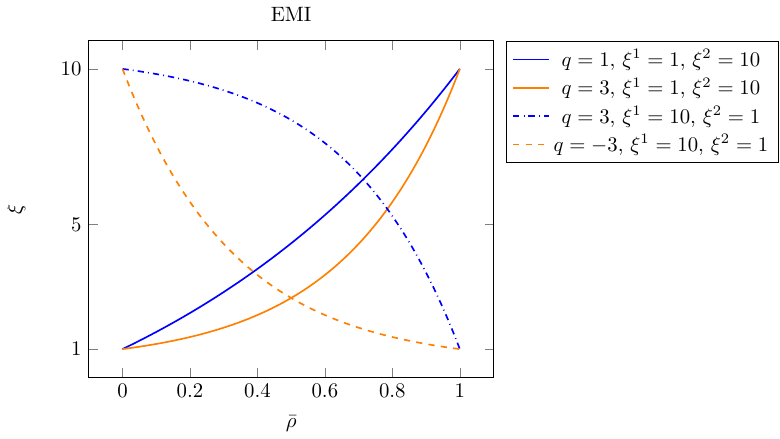}
	\caption{The EMI interpolation in \eqref{eq:EMI} with example values on $\xi^1$, $\xi^2$, and $q$.}
	\label{fig:MaterialInterpolation}
\end{figure}

To exemplify the EMI and the influence of the penalization parameter $q$, consider the interpolation between two materials
\begin{align}
\xi &= \xi^1 + \chi_q(\Bar{\rho})\qty(-\xi^1+\xi^2),
\label{eq:EMI}
\end{align}
which is shown in Fig. \ref{fig:MaterialInterpolation}. The sign of the parameter $q$ determines if the highest sensitivities are at $\Bar{\rho}=0$ or $\Bar{\rho}=1$. The interpolation is also symmetric around $\Bar{\rho}=0.5$ when switching the order of the phases and the sign of $q$.

The three phase multi-material interpolation can now be formulated as
\begin{equation}
\xi = \xi^{0} + \chi_{q_1^\xi}(\Bar{\rho}_1) \qty(-\xi^{0}+\xi^{1}+\chi_{q_2^\xi}(\Bar{\rho}_2)\qty(-\xi^{1}+\xi^{2})),
\label{eq:mmEMI}
\end{equation}
where $\xi^i$ represents the material parameters $K$, $G$, $\varepsilon_r$ and $c_e$, in the void material and the two solid materials, electrode and EAP. In this interpolation scheme, $\Bar{\rho}_1=0$ models void and $\Bar{\rho}_1=1$ solid material. Similarly for the solid material, where $\Bar{\rho}_2=0$ models electrode and $\Bar{\rho}_2=1$ models EAP. Due to the symmetry properties of EMI, the interpolation is independent of the order of the two solid materials.

\subsection{Sensitivity analysis}
\label{sec:sensitivity}

The design updates are obtained using the gradient based MMA, see \textcite{svanberg_method_1987}, which uses gradients of the objective and constraint functions $g_j$ w.r.t. each of the design variables $\vb*{\rho}_i$. Because regularization is performed via the PDE filter in \eqref{eq:Filter} and \eqref{eq:FilterBC}, the chain rule is applied to the objective and constraint functions such that
\begin{equation}
\pdv{g_j}{\vb*{\rho}_i}
= \pdv{g_j}{\vb*{\Tilde{\rho}}_i}\pdv{\vb*{\Tilde{\rho}}_i}{\vb*{\rho}_i},
\end{equation}
where $\pdv{\vb*{\Tilde{\rho}}_i}{\vb*{\rho}_i}$ is obtained from the discretized form of the filter in \eqref{eq:FilterFEM}.

The sensitivity of the objective function for maximizing displacements in \eqref{eq:g0maxDisp} is calculated with the adjoint method wherein the original objective function $g_0$ is augmented by an adjoint vector $\vb*{\mu}$ multiplied by the coupled residual in \eqref{eq:LinarisedSystem},
\begin{equation}
\Tilde{g}_0(\vb*{a},\vb*{\Tilde{\rho}}) = g_0 - \vb*{\mu}^T\vb*{r}.
\end{equation}
The sensitivities w.r.t. the filtered densities $\vb*{\Tilde{\rho}}_i$ is thus computed as
\begin{align}
\pdv{\Tilde{g}_0(\vb*{a},\vb*{\Tilde{\rho}})}{\vb*{\Tilde{\rho}}_i} 
&= \pdv{\Tilde{g}_0}{\vb*{\Tilde{\rho}}_i}
+ \pdv{\Tilde{g}_0}{\vb*{a}}\pdv{\vb*{a}}{\vb*{\Tilde{\rho}}_i} 
= - \vb*{\mu}^T\pdv{\vb*{r}}{\vb*{\Tilde{\rho}}_i} + \qty( \vb*{l}^T  - \vb*{\mu}^T\vb*{K} )\pdv{\vb*{a}}{\vb*{\Tilde{\rho}}_i},
\label{eq:g0augmented}
\end{align}
where the implicit sensitivity $\pdv{\vb*{a}}{\vb*{\Tilde{\rho}}_i}$ is annihilated by assigning $\vb*{\mu}$ to fulfill
\begin{equation}
\vb*{K}\vb*{\mu}= \vb*{l},
\label{eq:adjointProblem}
\end{equation}
where $\vb*{K}$ is the coupled stiffness matrix in \eqref{eq:LinarisedSystem}. Homogeneous Dirichlet boundary condition is enforced in all prescribed displacements and potential dofs.

Once \eqref{eq:adjointProblem} is solved for $\vb*{\mu}$, the sensitivity reduces to
\begin{align}
\pdv{\Tilde{g}_0(\vb*{a},\vb*{\Tilde{\rho}})}{\vb*{\Tilde{\rho}}_i}
&= 
- \vb*{\mu}^T\pdv{\vb*{r}}{\vb*{\Tilde{\rho}}_i}.
\label{eq:g0sens}
\end{align}

\section{Numerical examples}
\label{sec:NumericalEx}

The parameters for the material phases are given in Tab. \ref{tab:MatProp}. The EAP material is modeled with electro-mechanical coupling while the electrode material is uncoupled. To mimic the electrical conductivity of the electrode, a very large value of the $c_e$ parameter in \eqref{eq:AugmentedFreeEnergy} is used, while a low value is used for the EAP material. Vacuum permittivity is assumed for the void material and without any electro-mechanical coupling. The electrode is stiffer than EAP and the void stiffness is sufficiently small to not have a significant mechanical influence while avoiding numerical convergence issues. Additional assumptions are that $D_N=0$ in \eqref{eq:MatGaussLawBC} and $\vb*{T}_N=\vb*{0}$ in \eqref{eq:SpatTotalBalanceOfMomentumNeumannBC}.

\begin{table}[h]
	\centering
	\caption{Material parameters for the different phases}
	\begin{tabular}{l|llll}
		Material & $K$ [\si{\mega\pascal}] & $G$ [\si{\mega\pascal}] & $c_\text{e}$ [\si{\newton\per\square\volt}] & $\varepsilon_r$ [-] \\ \hline
		\\[-7pt]
		EAP & 0.6 & 0.1 & $-\frac{1}{2}\varepsilon_0$ & 4.7 \\[3pt]
		Electrode & $0.6\cdot10$ & $0.1\cdot10$ & $-\frac{1}{2}\varepsilon_0\cdot10^{5}$ & 0 \\[3pt]
		Void & $0.6\cdot10^{-9}$ & $0.1\cdot10^{-9}$ & $-\frac{1}{2}\varepsilon_0$ & 0 \\
	\end{tabular}
	\label{tab:MatProp}
\end{table}

To reduce the computational cost, the free space discretization is coarser than the solid body and graded with larger elements at large distances from the optimized structure. Algorithm \ref{alg:OptWithoutConnectivity} presents a pseudo-code for implementation.

\begin{algorithm}
	\caption{Optimization algorithm}
	\label{alg:OptWithoutConnectivity}
	\begin{algorithmic}
		\State Set initial values on all parameters
		\State Set $\vb*{\rho}_i$ for the initial design
		\While{$\abs{g_0^k-g_0^{k-1}}>TOL\abs{g_0^{k}}$}
		\State Update relevant parameters
		\State Apply filter \eqref{eq:FilterFEM} to obtain $\vb*{\Tilde{\rho}}_i$ from $\vb*{\rho}_i$
		\State Apply threshold \eqref{eq:Hproj} to obtain $\vb*{\Bar{\rho}}_i$ from $\vb*{\Tilde{\rho}}_i$ 
		\State Solve state problem \eqref{eq:TotalLagrangianFEresidualsPot} and \eqref{eq:TotalLagrangianFEresidualsDef} using \eqref{eq:LinarisedSystem}
		\State Calculate objective and constraints $g_{j}$, $j=0,1,...,n_{constr}$
		\State Compute sensitivities using the adjoint method
		\State Update element design variables $\vb*{\rho}_i$ using MMA
		\EndWhile
	\end{algorithmic}
\end{algorithm}

\begin{figure}
	\centering
	\includegraphics{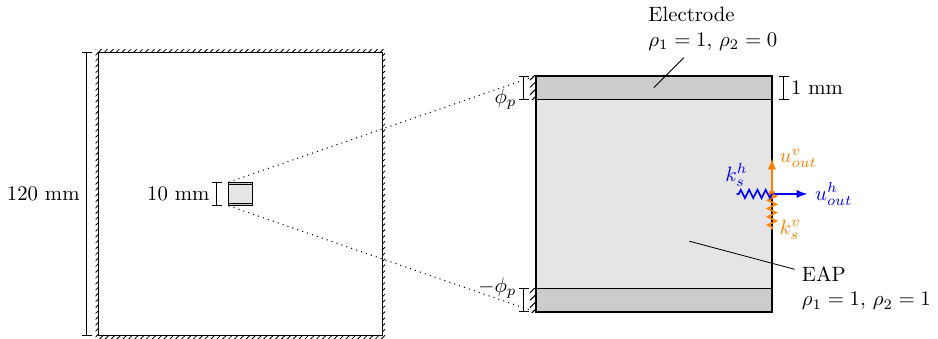}
	\caption{Illustration of the geometry and boundary conditions for the actuator with thickness 1 mm. The square design domain is surrounded by a large free space. The initial design and the two cases for the optimization are presented.}
	\label{fig:exampleActuator}
\end{figure}

An actuator suspended in a free space, as illustrated in Fig. \ref{fig:exampleActuator}, is activated by a prescribed positive and negative potential in the upper and lower left edges respectively, where $\phi_p=3$ kV, i.e., the external source is connected to these two electrode locations. It is assumed that the two electrode locations are fixed in space, i.e., clamped. The two objective functions investigated in this example are vertical and horizontal output displacements, $u_{out}^v$ and $u_{out}^h$, respectively, also illustrated in Fig. \ref{fig:exampleActuator}. Two springs with spring stiffness $k_s^v=k_s^h=10^{-3}d G^{\text{EAP}}$ are connected to the corresponding output ports to simulate the stiffness of the actuated object and with the thickness $d=1\text{ mm}$. The vector $\vb*{l}$ in \eqref{eq:g0maxDisp} is defined such that only the displacement $u_{out}^v$ or $u_{out}^h$ on the right port indicated in Fig. \ref{fig:exampleActuator} are extracted from $\vb*{a}$.

The initial design in Fig. \ref{fig:exampleActuator} consists of electrode and EAP material indicated with darker and lighter gray, respectively. The initial density values are $\rho_1=1$ everywhere and $\rho_2=0$ for the electrode material and $\rho_2=1$ for the EAP material.

The design domain $\mathcal{B}_{0}$ is discretized using 200x200x1=40 000 8-node brick elements while the free space $\mathcal{F}_0$ is resolved using an 11 285 element graded mesh. Plane strain is enforced by constraining the displacement in the out-of-plane direction. Although 2D elements could be used, a 3D framework is used as this allows for a smooth transition to 3D in future works.

The filter parameter in \eqref{eq:Filter} is $l_i=\SI{0.25}{\milli\meter}$. A continuation scheme is employed for the thresholding parameter $\beta$ in \eqref{eq:Hproj}, which is initiated to $\beta=1$ and then increased by 20\% every 10th iteration to a terminal value of 20. The parameter $\eta=0.5$ is fixed.

Inspired by \textcite{ortigosa_density-based_2021}, the electro-mechanical coupling terms are penalized more than the purely mechanical to ensure low electro-mechanical coupling in void and intermediate regions. The electrical terms has the same penalization as the electro-mechanical. A continuation scheme is utilized where the initial and final penalization parameters for EMI are given in Tab. \ref{tab:PenalisationParameters} and the values is updated with 20\% every 10th iteration until the final values are reached.

\begin{table}[h]
	\centering
	\caption{Initial and final penalization parameters for EMI}
	\begin{tabular}{c|ccc|ccc}
		& \multicolumn{3}{c|}{Initial} & \multicolumn{3}{c}{Final} \\
		$i$ & $q_i^{m}$  & $q_i^{mel}$  & $q_i^{el}$ & $q_i^{m}$  & $q_i^{mel}$  & $q_i^{el}$ \\ \hline
		1 & 1 & 2 & 2 & 4 & 8 & 8\\
		2 & -1 & 2 & -2 & -4 & 8 & -8
	\end{tabular}
	\label{tab:PenalisationParameters}
\end{table}

The output displacement varies with multiple orders of magnitude, therefore is $g_0$ is replaced by
\begin{equation}
\Hat{g}_0 = a_1\text{asinh}\qty(a_2 g_0),
\label{eq:g0asinh}
\end{equation}
where $a_1=10$ and $a_2=10^7$ to obtain a suitable MMA scaling. The allowed volume fractions are $\alpha_1=0.2$ and $\alpha_2=0.1$ and both constraint functions $g_1$ and $g_2$ are scaled by 10. Design updates continues until $\abs{g_0^{k}-g_0^{k-1}}<TOL\abs{g_0^{k}}$ during three consecutive iterations and all constraints were fulfilled, with the relative tolerance $TOL=10^{-4}$.

A continuation scheme is employed for $\alpha$ and $a_d$ in \eqref{eq:addCRISPterms}, while $\delta=10^{-9}$ is constant. The penalization in \eqref{eq:CRISP} is initiated to $a_d=0$ and after iteration 200 and until iteration 300, $a_d$ is updated every 20th iteration such that $a_d=\abs{\hat{g}_0}$. The parameter $\alpha$ is initiated as $\alpha=0.9$ and after iteration 200 updated every 10th iteration with $\alpha=\text{max}(\alpha-0.1,0.05)$.

The optimization formulation in \eqref{eq:SO} now becomes
\begin{equation}
(\mathbb{TO})_1
\begin{cases}
\displaystyle \min_{\vb*{\rho}_1,\vb*{\rho}_2} \bar{g}_0 = \hat{g}_0\qty(g_0) + a_{d} \sum_{i=1}^2 d_\delta^\alpha(\Bar{\rho}_i)\\
\displaystyle \text{subjected to}
\begin{cases}
\displaystyle g_j = \frac{V_j\qty(\Bar{\rho}_j)}{\alpha_j V_{DD}}-1\leq0, \quad j=1,2\\
\displaystyle 0\leq \rho_i^e\leq 1,\quad \forall e\in[1,...,n_{elm}], i\in[1,2]
\end{cases}
\end{cases}
\label{eq:P1}
\end{equation}
Default MMA parameters is used with the exceptions for $\texttt{GHINIT}=0.2$, $\texttt{GHINCR}=1.1$, $\texttt{GMAXJ = XVAL(J)-0.0001*XMMJ}$ and $\texttt{HMINJ = XVAL(J)+0.0001*XMMJ}$.

\begin{figure*}
	\centering
	\begin{subfigure}[b]{0.48\textwidth}
		\centering
		\includegraphics[width=\textwidth]{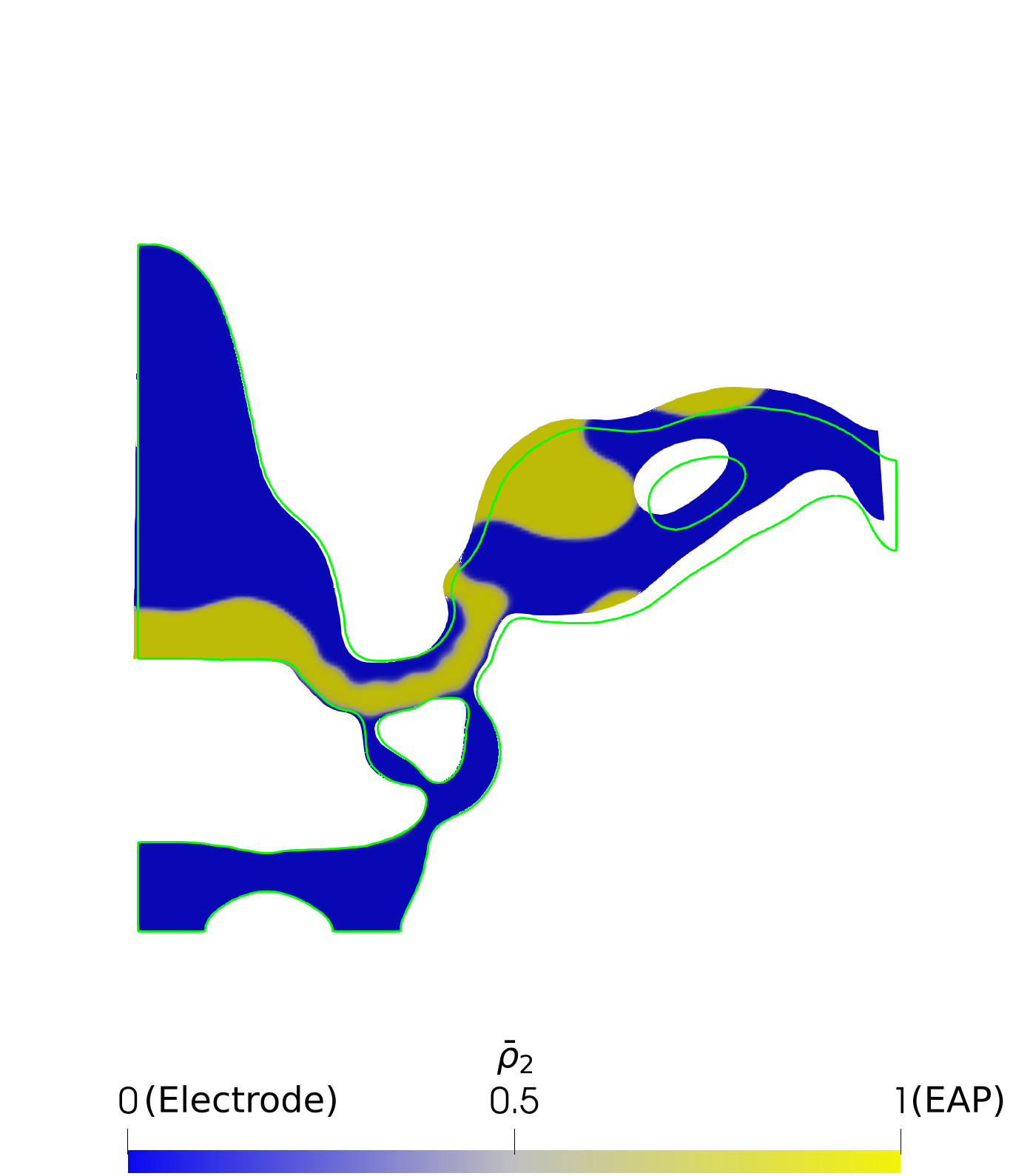}
		\caption{}
		\label{fig:OptimizedActuators1A}
	\end{subfigure}
	\hfill
	\begin{subfigure}[b]{0.48\textwidth}
		\centering
		\includegraphics[width=\textwidth]{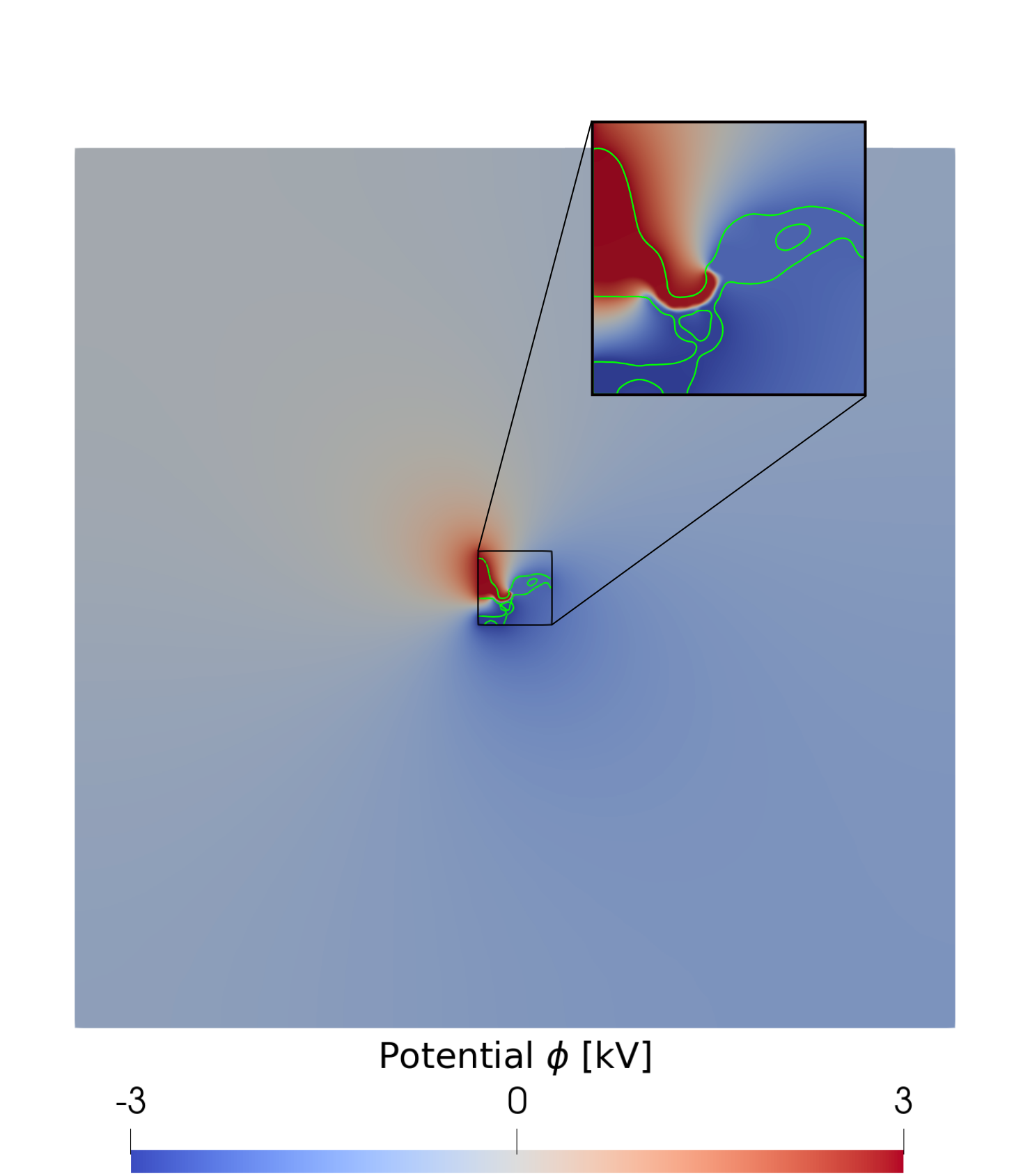}
		\caption{}
		\label{fig:OptimizedActuators1B}
	\end{subfigure}
	\hfill
	\begin{subfigure}[b]{0.48\textwidth}
		\centering
		\includegraphics[width=\textwidth]{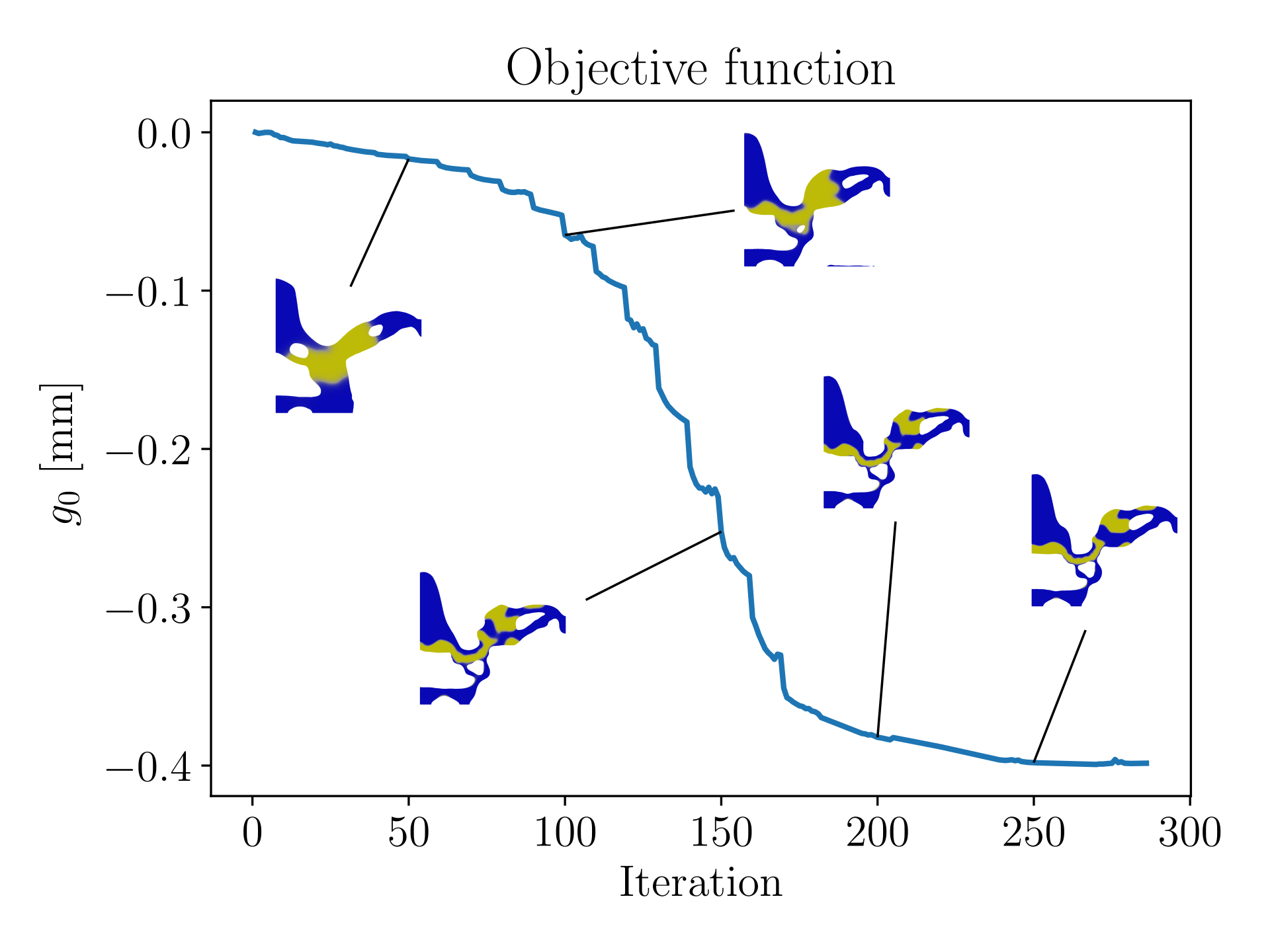}
		\caption{}
		\label{fig:OptimizedActuators1C}
	\end{subfigure}
	\hfill
	\begin{subfigure}[b]{0.48\textwidth}
		\centering
		\includegraphics[width=\textwidth]{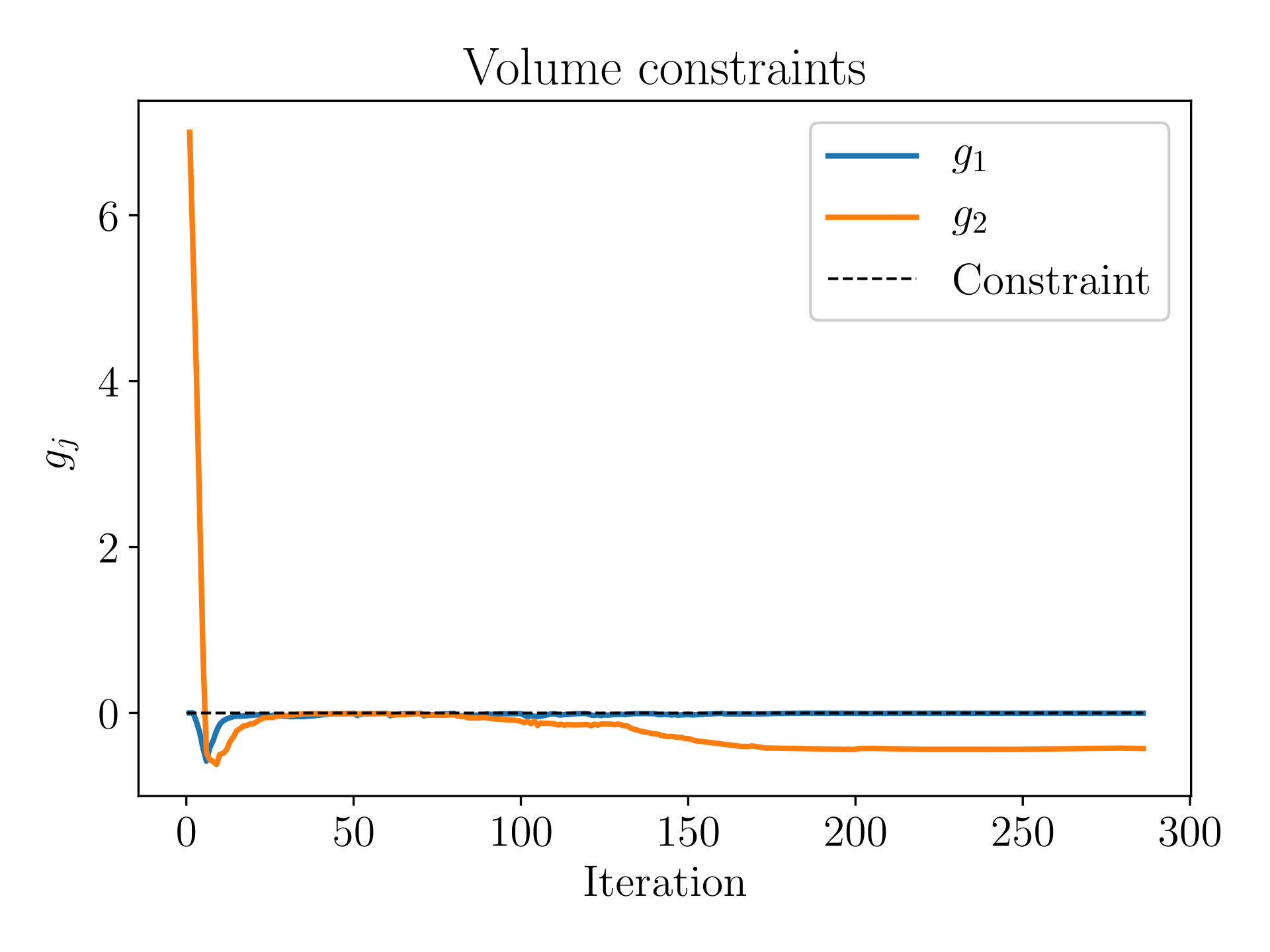}
		\caption{}
		\label{fig:OptimizedActuators1D}
	\end{subfigure}
	\caption{Optimized design for maximizing vertical displacement $u_{out}^{v}$. a) Deformed structure cut-out where $0.5\leq \Bar{\rho}_1$ and where $\Bar{\rho}_2=0$ indicates electrode material and $\Bar{\rho}_2=1$ EAP. Solid green lines indicates the undeformed configuration. b) Electric potential $\phi$ in the undeformed total domain with the design domain highlighted as a black square and outline of the final design in green. Note that the highest gradient is concentrated to the EAP material. c) Objective function with converged value $g_0=-0.399$ mm at iteration 286. The designs for selected iterations are also shown. d) Volume constraints.}
	\label{fig:OptimizedActuators1}
\end{figure*}

Fig. \ref{fig:OptimizedActuators1A} shows the optimized design $0.5\leq\Bar{\rho}_1$ when the vertical displacement is maximized. Electrode material is indicated with $\Bar{\rho}_2=0$ and EAP with $\Bar{\rho}_2=1$. Some electrode material is continuously connected from the electrical sources in the corners to opposite sides of the EAP material. Due to this, the electric potential gradient, and thereby the electric field, is concentrated to within the EAP material, see Fig. \ref{fig:OptimizedActuators1B}. This results in a thin active EAP layer which causes the deformation. Since the EAP material is curved with thicker electrode material on the upper side, it bends during activation. Traditional EAP structures are constructed in a similar fashion, where a thin layer of EAP material is sandwiched between electrodes.

From \ref{fig:OptimizedActuators1B}, it is obvious that the highest electric fields occur within or close to the design domain. At the boundary $\partial\mathcal{F}_0^\infty$ the electric field is zero due to the boundary condition in \eqref{eq:MatGaussLawBCinfty} and the potential approaches zero, which is expected.

Since the electrode material is stiffer than the EAP, it also stiffens the connection between the active layer and the output port. As the maximum amount of electrode material is utilized, see Fig. \ref{fig:OptimizedActuators1D}, additional EAP material is used to stiffen the structure, but not activated.

\begin{figure*}
	\centering
	\begin{subfigure}[b]{0.48\textwidth}
		\centering
		\includegraphics[width=\textwidth]{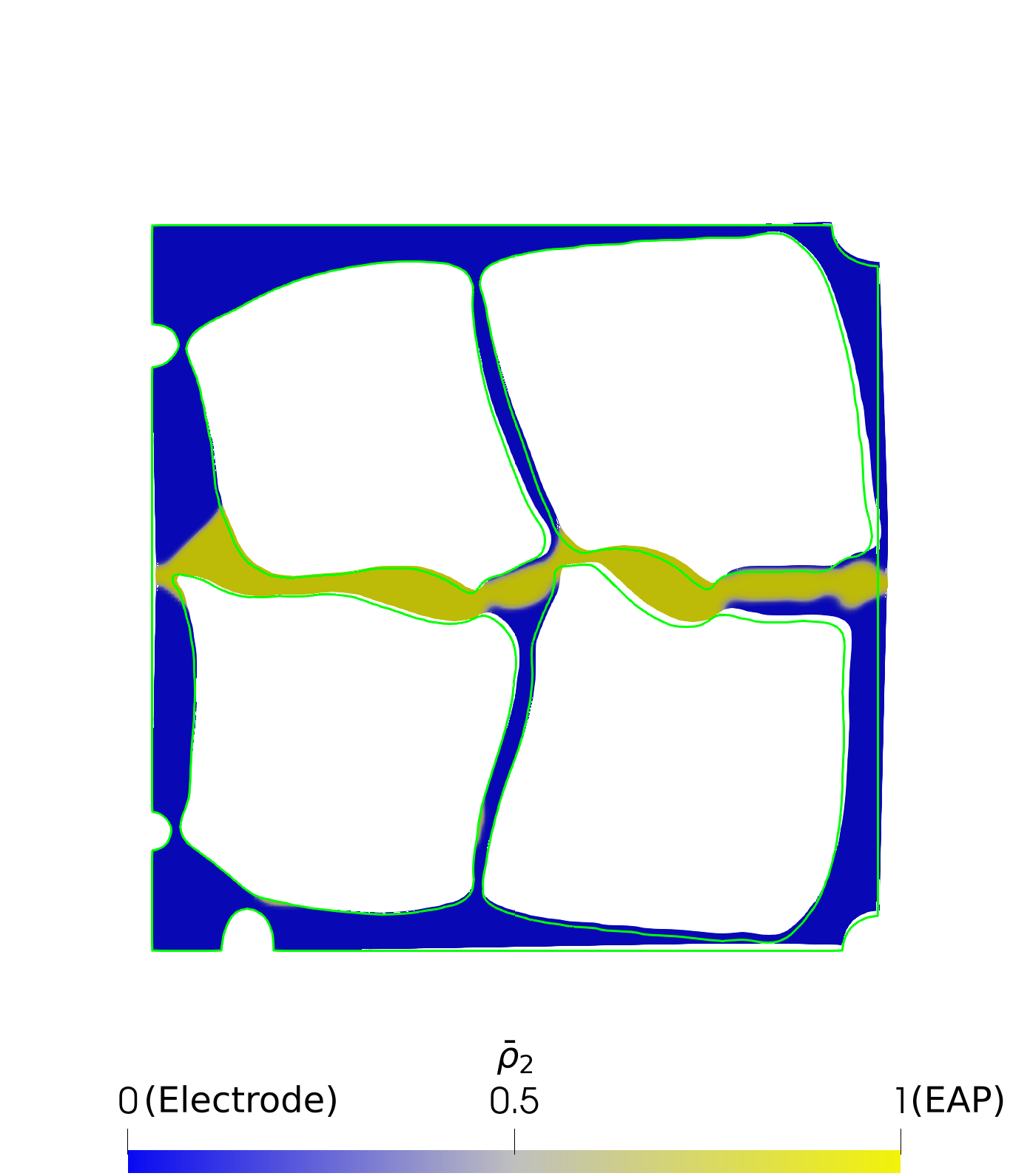}
		\caption{}
		\label{fig:OptimizedActuators2A}
	\end{subfigure}
	\hfill
	\begin{subfigure}[b]{0.48\textwidth}
		\centering
		\includegraphics[width=\textwidth]{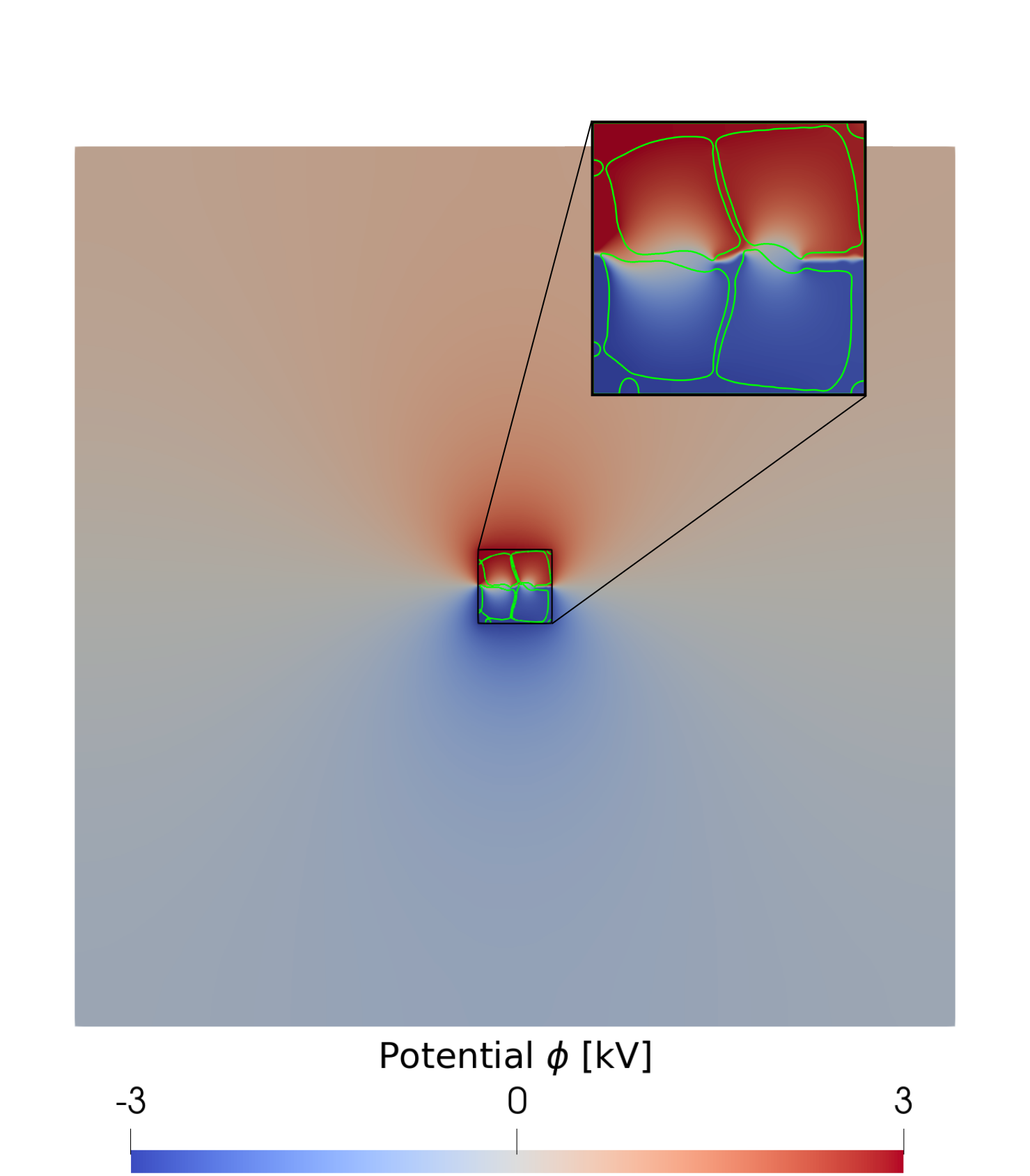}
		\caption{}
		\label{fig:OptimizedActuators2B}
	\end{subfigure}
	\hfill
	\begin{subfigure}[b]{0.48\textwidth}
		\centering
		\includegraphics[width=\textwidth]{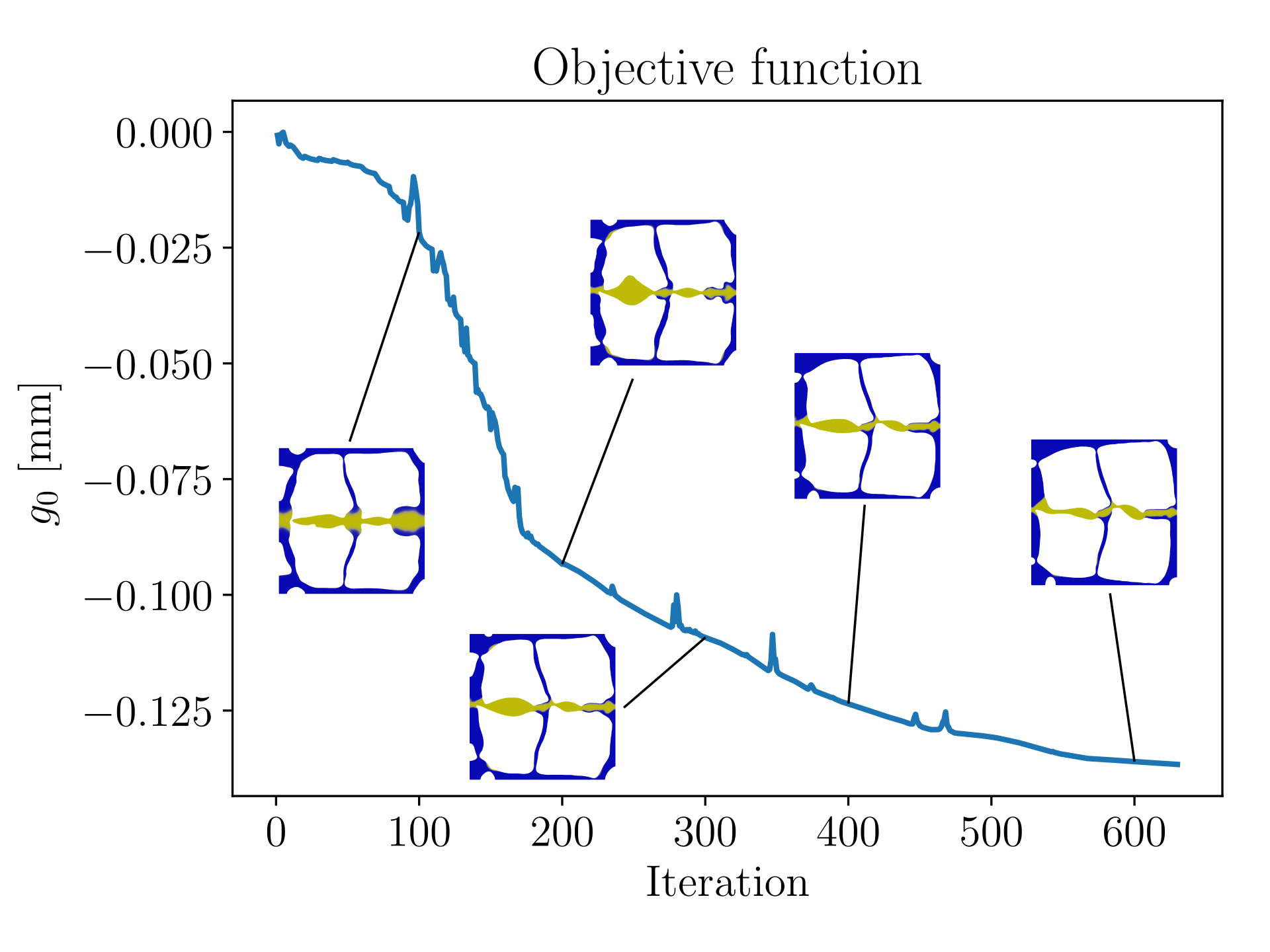}
		\caption{}
		\label{fig:OptimizedActuators2C}
	\end{subfigure}
	\hfill
	\begin{subfigure}[b]{0.48\textwidth}
		\centering
		\includegraphics[width=\textwidth]{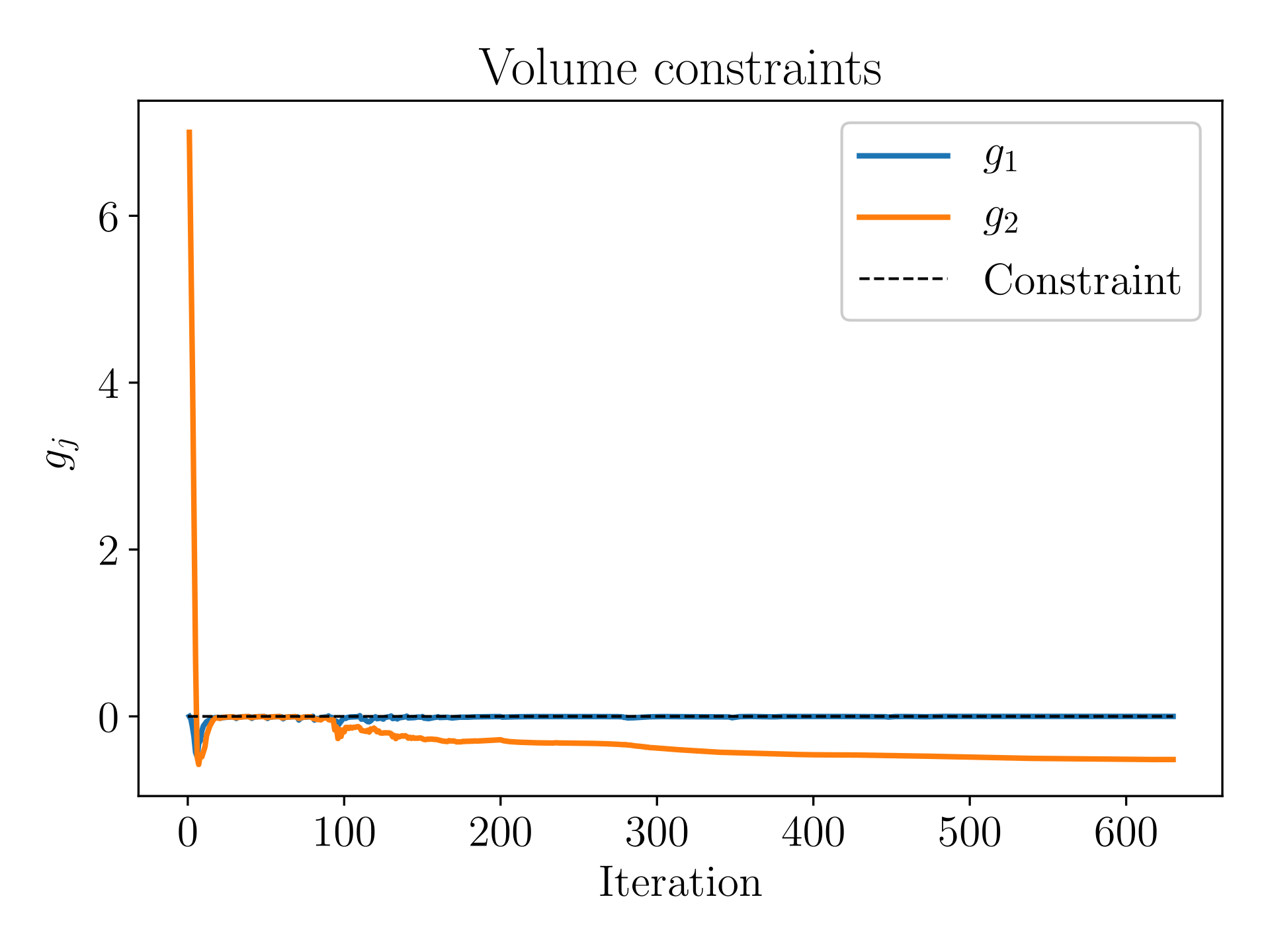}
		\caption{}
		\label{fig:OptimizedActuators2D}
	\end{subfigure}
	\caption{Optimized design for maximizing horizontal displacement $u_{out}^{h}$. a) Deformed structure cut-out where $0.5\leq \Bar{\rho}_1$ and where $\Bar{\rho}_2=0$ indicates electrode material and $\Bar{\rho}_2=1$ EAP. Solid green lines indicates the undeformed configuration. b) Electric potential $\phi$ in the undeformed total domain with the design domain highlighted as a black square and outline of the final design in green. Note that the highest gradient is concentrated to the EAP material. c) Objective function with converged value $g_0=-0.137$ mm at iteration 630. The designs for selected iterations are also shown. d) Volume constraints.}
	\label{fig:OptimizedActuators2}
\end{figure*}

Fig. \ref{fig:OptimizedActuators2A} shows the optimized design when maximizing the horizontal displacement. Also in this case, the electrode material is continuously connected from the electrical sources in the corners to opposite sides of a thin EAP layer. The EAP layer stretches through the whole structure where the electric field is concentrated to three different parts, as can be seen in Fig. \ref{fig:OptimizedActuators2B}. The generated deformations is thus transferred through the EAP layer to the output port, creating a horizontal dominated deformation.

The potentials in Figs. \ref{fig:OptimizedActuators2B} and \ref{fig:OptimizedActuators1B} are similar. The electrodes are shaped differently and much longer, which influences the free space potential, while still fulfilling the boundary condition on the far-field boundary.

Comparing Figs. \ref{fig:OptimizedActuators1C} and \ref{fig:OptimizedActuators2C} shows that the latter case requires more iterations to converge. During the first 200 iterations they behave similarly, but then the convergence is much slower when the horizontal displacement is optimized. The constraint functions in Figs. \ref{fig:OptimizedActuators1D} and \ref{fig:OptimizedActuators2D} behaves similarly, where the electrode material fulfills the volume fraction while the EAP volume constraint in inactive.

\section{Conclusions}
\label{sec:Conclusions}

This work presented a topology optimization methodology for generating EAP structures with arbitrary electrode and EAP layouts by utilizing density-based, multi-material topology optimization. Numerical test showed that common interpolation schemes were not able to create distinct material phases for the current problem. Due to this, an interpolation based on exponential functions, denoted exponential material interpolation, was introduced which allows for a large flexibility to control the material interpolation.

Standard regularization using the PDE filter and the smooth Heaviside projection was utilized on each density field. Despite the smooth Heaviside projection, the optimization might generate designs with intermediate densities. A recent work, see \textcite{thillaithevan_inverse_2024}, suggested a penalization that were able to further decrease the amount of intermediate densities. This was successfully adopted in the current work by penalizing both density fields separately.

Most research on EAPs neglects the influence of the electric field in the free space surrounding a solid body. To model this influence, a truncated, extended domain method with vacuum permittivity and very low stiffness was used in this work. This is included in the governing equations for non-linear electro-elasticity by considering the computational domain as both the solid body and the free space with boundary conditions on the shared boundary.

The presented examples illustrates that this methodology is able to generate designs where the electrode and EAP material can take arbitrary layouts and still be connected to the electrical sources. Additionally, the generated designs consist of thin layers of EAP material with electrodes on each side, concentrating the electric field to the EAP material. This is similar to how traditional EAP structures are constructed, where a thin EAP layer is sandwiched between compliant electrodes.

\section*{Acknowledgements}
The Swedish Research Council, Grants No. 2021-03851 and 2020-04364, is gratefully acknowledged for providing funding. Additionally, the authors would like to thank Krister Svanberg for his MMA implementation in Fortran.

\section*{Author contributions: CRediT}
\textbf{Daniel Hård:} Conceptualization, Data curation, Formal Analysis, Investigation, Methodology, Software, Validation, Visualization, Writing – original draft, Writing – review \& editing. \textbf{Mathias Wallin:} Funding acquisition, Supervision, Writing – review \& editing. \textbf{Matti Ristinmaa:} Funding acquisition, Supervision, Writing – review \& editing.

\section*{Funding data}
This work was supported by the Swedish Research Council, Grants No. 2021-03851 and 2020-04364.

\section*{Conflict of interest}
On behalf of all authors, the corresponding author states that there is no conflict of interest.

\printbibliography

\end{document}